\documentclass[runningheads]{llncs}
\usepackage[T1]{fontenc}
\usepackage{subfig} 
\usepackage{multirow}
\usepackage{xcolor} 
\usepackage{graphicx}
\usepackage{amsmath}
\usepackage{amssymb}
\usepackage{algorithm}
\usepackage{algpseudocode}
\usepackage{adjustbox} 
\usepackage{caption}
\usepackage{hyperref}
\usepackage{url}
\usepackage{soul}

\makeatletter
\newcommand{\printfnsymbol}[1]{%
  \textsuperscript{\@fnsymbol{#1}}%
}
\makeatother

\begin{document}
\title{Conditional diffusion model with spatial attention and latent embedding for medical image segmentation}
\titlerunning{cDAL for Medical Image Segmentation}
%
\author{Behzad Hejrati\inst{1} \thanks{Equal contribution}  
\and
Soumyanil Banerjee \inst{1} \printfnsymbol{1} 
\and
Carri Glide-Hurst\inst{2} 
\and
Ming Dong\inst{1}
\thanks{Corresponding author}
}
%
\institute{Department of Computer Science, Wayne State University, MI, USA 
\email{\{b.hejrati,s.banerjee,mdong\}@wayne.edu} \and
Department of Human Oncology, University of Wisconsin-Madison, WI, USA \\
\email{glidehurst@humonc.wisc.edu}}
%
\authorrunning{B. Hejrati, S. Banerjee et. al.}
%
%
\maketitle              
\begin{abstract}
Diffusion models have been used extensively for high quality image and video generation tasks. In this paper, we propose a novel conditional diffusion model with spatial attention and latent embedding (cDAL) for medical image segmentation. In cDAL, a convolutional neural network (CNN) based discriminator is used at every time-step of the diffusion process to distinguish between the generated labels and the real ones. A spatial attention map is computed based on the features learned by the discriminator to help cDAL generate more accurate segmentation of discriminative regions in an input image. Additionally, we incorporated a random latent embedding into each layer of our model to significantly reduce the number of training and sampling time-steps, thereby making it much faster than other diffusion models for image segmentation. We applied cDAL on 3 publicly available medical image segmentation datasets (MoNuSeg, Chest X-ray and Hippocampus) and observed significant qualitative and quantitative improvements with higher Dice scores and mIoU over the state-of-the-art algorithms. The source code is publicly available at \url{https://github.com/Hejrati/cDAL/}.

\end{abstract}

\section{Introduction}
\label{sec:intro}
Medical image segmentation is a crucial task in clinical practice with applications including disease diagnosis, radiotherapy and surgical treatment planning \cite{wang2022medical}\cite{masood2015survey}. A major challenge in the segmentation process are the manual annotations performed by a trained clinician, which is a time-consuming process that is not scalable. Hence, automation of the segmentation with deep learning algorithms have been a key area of research for the last several years. The algorithms which produced state-of-the-art results for end-to-end 2D and 3D medical image segmentation task include the U-Net \cite{ronneberger2015u} and the 3D U-Net \cite{cciccek20163d}, respectively. 

Diffusion models are a class of generative models where a neural network is trained to remove the noise from an image which was produced during the forward process with a pre-defined noise schedule. This trained neural network is then used in the sampling process to iteratively remove the Gaussian noise from an image and eventually generate high quality samples by starting from pure Gaussian noise \cite{ho2020denoising}\cite{sohl2015deep}\cite{nichol2021improved}. Diffusion models generate more diverse images than Generative Adversarial Networks (GANs) and have recently outperformed GANs for the generation of high resolution images \cite{rombach2022high}\cite{dhariwal2021diffusion}.

Image segmentation with diffusion models is a challenging task due to the deterministic nature of image segmentation as opposed to the stochastic nature of diffusion models. Hence, diffusion models have been used for supervised medical image segmentation tasks to model the distribution of labels resulting from independent annotators of the same image \cite{rahman2023ambiguous}\cite{amit2023annotator}. When the segmentation labels are scarce, the semantic representation from intermediate layers of a pretrained diffusion model is used to train a simple pixel-level classifier with the small set of available labels \cite{baranchuk2021label}. In \cite{wolleb2022diffusion}\cite{amit2021segdiff}, the image is used as a condition to a diffusion model during the label generation process, which is repeated a few times due to the stochastic nature of diffusion models. The mean of all such label generations is considered as the final segmentation map.

Diffusion models have a common drawback that the sampling procedure to generate the images from pure Gaussian noise is a time-consuming process. This problem was addressed with several interesting ideas such as non-markovian diffusion process \cite{song2020denoising} and distillation in diffusion models \cite{salimans2022progressive}\cite{song2023consistency}. But, faster sampling typically results in degradation of the generated image quality. Hence, there was a need to tackle the generative learning trilemma of achieving fast sampling, higher quality and diversified image samples. This trilemma was addressed with the denoising diffusion GANs \cite{xiao2021tackling} by modeling the denoising distribution with a complex multimodal distribution. 

In this work, we propose a novel method of using a conditional diffusion model with spatial attention and latent embedding (cDAL) for medical image segmentation. During training, cDAL uses a diffusion model to predict the unperturbed segmentation labels from a noisy label. The image is encoded and passed as a condition to the input of the label diffusion model. During each diffusion time-step, we incorporate a separate discriminator to distinguish between the ground-truth labels and the generated ones. We use the spatial attention map learned by the discriminator \cite{emami2020spa} to compute attention-based labels as the input of the diffusion model so that it can focus on these discriminative regions during segmentation. We also incorporate a random latent embedding into each layer of the diffusion model to reduce the number of diffusion time-steps in both training and sampling. 

Our main contributions are: (i) We incorporated a separate discriminator for each diffusion time-step and guided the diffusion process with the spatial attention map learned from the discriminator. (ii) We used a random latent embedding for each layer of the diffusion model which helped in reducing both the training and sampling time-steps by modeling the denoising distribution with a complex multimodal distribution. (iii) We performed extensive experiments on two 2D binary (MoNuSeg and chest X-ray) and one 3D multi-class (Hippocampus) public medical image segmentation datasets and observed significant quantitative and qualitative improvements over state-of-the-art methods.    

\section{Method}
\label{sec:method}

In the following sections, we provide a detailed description of each component of our proposed cDAL architecture as shown in Fig. \ref{fig:figure1}.

\begin{figure*}[ht]
\centering
\includegraphics[width=0.99\textwidth]{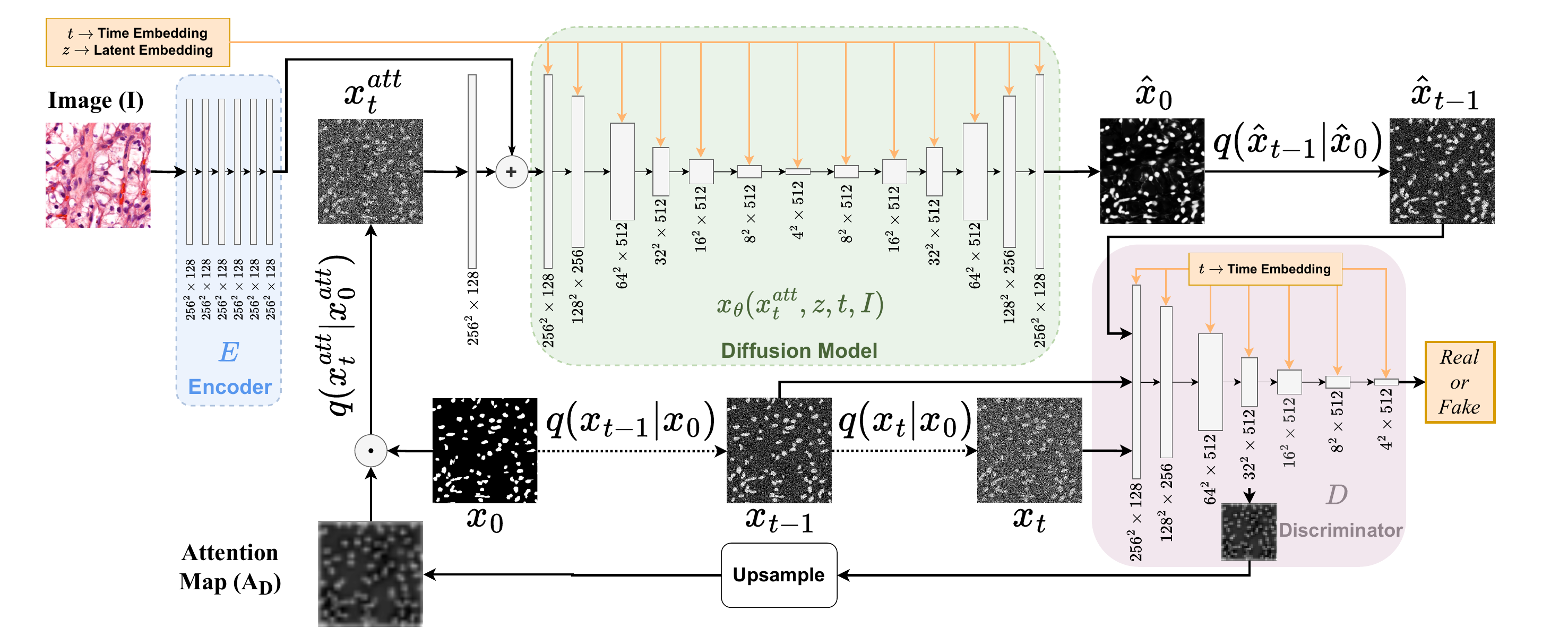}
\caption{Our proposed conditional diffusion model with spatial attention and latent embedding (cDAL) for medical image segmentation.}
\label{fig:figure1}
\end{figure*}

\subsection{Conditional diffusion model for image segmentation}
\label{subsec: cDDPM}

There is inherent ambiguity in medical image segmentation as the delineation of the same image differs among experts. 
In our proposed cDAL, we utilized the stochastic nature of DDPM to approximate this process and generate multiple predictions during inference. Subsequently, we take the mean of the predictions and threshold them to obtain more accurate segmentation masks compared to deterministic models such as U-Net.

DDPM \cite{ho2020denoising} consists of a markov-chain forward process where Gaussian noise is gradually added to perturb the data distribution in $T$ time-steps. The forward process $q$ is given by the joint distribution: $q(x_{1:T} | x_{0}) = \prod_{t=1}^{T} q(x_{t} | x_{t-1})$, where for each step $t$, the forward process is: $q(x_t | x_{t-1}) = \mathcal{N}(x_t; \sqrt{1-\beta_t} x_{t-1}, \beta_t I)$. Here, $x_0$ is sampled from the data distribution, $T$ is the number of time-steps, $\beta_t$ is the predefined noise schedule, $\mathcal{N}$ denotes the Gaussian distribution and $I$ is a $n \times n$ shaped identity matrix of the same shape as the data $x_0$. The cumulative process from $x_0$ to $x_t$ is represented as: $x_t = \sqrt{\bar{\alpha}_t} x_0 + (1-\bar{\alpha}_t) \epsilon, \epsilon \sim \mathcal{N}(0, I_{n \times n})$. Here, $\bar{\alpha}_t = \prod_{s=1}^{t} (1-\beta_s)$ is the cumulative scaling factor used during the forward process $q(x_t | x_0) = \mathcal{N}(x_t; \sqrt{\bar{\alpha}_t} x_0, (1-\bar{\alpha}_t) I)$ to obtain sample $x_t$ at arbitrary time-step $t$.

The reverse process of DDPM to iteratively denoise the latent variables ($x_1, ..., x_T$) is parameterized by the joint distribution $p_{\theta}(x_{0:T})$ and given by:

\begin{equation}
\small
p_{\theta}(x_{0:T}) = p(x_T) \prod_{t=1}^{T} p_{\theta}(x_{t-1} | x_t) = p(x_T) \prod_{t=1}^{T} \mathcal{N}(x_{t-1}; \mu_{\theta}(x_t,t), \sigma_{t}^{2} I)
\label{equation:p_theta}
\end{equation}
where, $\mu_{\theta}(x_t,t)$, $\sigma_{t}^{2}$ and $\theta$ denote the mean, variance and parameters of the denoising model $p_{\theta}(x_{t-1} | x_t)$ respectively. 

By maximizing the evidence lower bound \cite{ho2020denoising}, we have the training loss function: $\arg \min_{\theta} \mathbb{E}_{x_0, \epsilon, t}[||\epsilon - \epsilon_{\theta}(x_t,t)||^2]$, where $\epsilon \sim \mathcal{N}(0, I_{n \times n})$ denotes pure Gaussian noise and $\epsilon_{\theta}$ denotes the predicted noise by the denoising network. During the sampling stage, the trained model $\epsilon_{\theta}$ is used to iteratively denoise the data, i.e. generate $x_{t-1}$ from $x_t$ for $t = T, T-1, ..., 1$ and eventually generate the data $x_0$ by starting from pure Gaussian noise $x_T \sim \mathcal{N}(0, I_{n \times n})$. 

This unconditional generation process of DDPM is suitable for image generation tasks where the goal is to model a data distribution. For image segmentation tasks, there exists an image and label pair $(I,x)$, where $I$ denotes the image and $x$ denotes the corresponding ground-truth label. Hence, for image segmentation tasks, diffusion models are helpful to generate a distribution of labels but it needs to have the image as a condition to generate relevant labels. 

In cDAL, we use the diffusion model as a generator $x_{\theta}$ with the image $I$ as a condition to guide the diffusion model to generate the label ($x$) corresponding to the image $I$ as shown in Fig. \ref{fig:figure1}. In our approach, instead of predicting the noise with the diffusion model $\epsilon_{\theta}$, we use the formulation provided by \cite{benny2022dynamic} and directly predict the clean label $\hat{x}_{0}$ using our diffusion model conditioned on the image, i.e. $\hat{x}_{0} = x_{\theta}(x_t, t, I)$, where $t$ is the time embedding.

\subsection{cDAL: spatial attention maps}
\label{subsec: cDDPM_attn_maps}
In the cDAL architecture, we incorporate a distinct CNN-based discriminator $D$ as shown in Fig. \ref{fig:figure1}. This discriminator is trained to differentiate between the ground-truth segmentation labels and the labels generated using our diffusion model $x_{\theta}(x_t, t, I)$.

More specifically, first the conditional diffusion model is frozen. The perturbed label $x_{t-1}$ is generated using the forward process and ground-truth labels, i.e. $x_{t-1} := q(x_{t-1}|x_0)$. Then, the discriminator $D$ uses $x_t := q(x_t|x_{t-1})$, $x_{t-1}$ and time-step $t$ as inputs to predict the label as real. The cross-entropy loss is used to update $D$. Subsequently, with the diffusion model still frozen, the output of the diffusion model is: $\hat{x}_{0} = x_{\theta}(x_t, t, I)$. With $\hat{x}_0$, $\hat{x}_{t-1}$ is sampled using the posterior distribution $q(\hat{x}_{t-1} | x_t, \hat{x}_0)$. Then, the discriminator uses $\hat{x}_{t-1}$, $x_t$ and $t$ as inputs to predict the label as fake (class 0), and the cross-entropy loss is used to update $D$ again.

Clearly, the discriminator $D$ learns the most discriminative features to differentiate between the real $x_{t-1}$ and predicted $\hat{x}_{t-1}$. Here, we use the feature maps of $D$ to generate the spatial attention map $A_D = \frac{1}{C}\sum_{i=1}^{C}F_i$, where, $F_i$ denotes the $i^{th}$ feature map of $D$ with $C$ channels.

The attention map $A_D$ highlights the spatial regions in the labels which are essential for our model to generate labels $\hat{x}_0$ that are close to the ground-truth $x_0$. We upsample the attention map $A_D$ to match the shape of ground-truth label $x_0$ and then perform element-wise multiplication with $x_0$ to get $x_{0}^{att} = x_0 \odot A_D$, where $\odot$ represents the Hadamard product. Subsequently, the forward process is used to transform $x_{0}^{att}$ to $x_{t}^{att}$ using $q(x_{t}^{att}|x_{0}^{att})$. The perturbed $x_{t}^{att}$ is fed to the conditional diffusion model to predict $\hat{x}_0$ as depicted in Fig. \ref{fig:figure1}. With discriminator $D$ fixed, the diffusion model loss is $||x_0 - x_{\theta}(x_{t}^{att}, t, I)||^2$, where, $x_0$ is the ground-truth label, $x_{\theta}$ is the denoising model dependent on the attention incorporated $x_{t}^{att}$. This loss is used to update the parameters $\theta$ of the conditional diffusion model.

\subsection{cDAL: Latent embedding}
\label{subsec: cDDPM_latent}

DDPM \cite{ho2020denoising} typically uses a large number of time-steps for both training and sampling since they use small step-sizes. Hence, the true denoising distribution is closer to a Gaussian distribution. When the denoising step size becomes larger, the denoising distribution deviates from a Gaussian and becomes a complex multi-modal distribution \cite{xiao2021tackling}. In cDAL, we use larger step sizes to perturb the label data $x_0 \sim q(x_0)$ in $T$ time-steps ($T \leq 4$) using the forward process $q(x_t | x_{t-1}) = \mathcal{N}(x_t; \sqrt{1-\beta_t} x_{t-1}, \beta_t I)$ with large variance $\beta_t$ in each time step. For the reverse process, the denoising model is given by:

\begin{equation}
\small
p_{\theta}(\hat{x}_{t-1} | x_t) := q(\hat{x}_{t-1} | x_t, \hat{x}_0 = x_{\theta}(x_{t}^{att}, t, I))
\label{equation:latent_emb}
\end{equation}
where, $\hat{x}_0$ is first predicted using $x_{\theta}(x_{t}^{att}, t, I)$ and then from the posterior distribution $q(\hat{x}_{t-1} | x_t, \hat{x}_0)$, $\hat{x}_{t-1}$ is sampled as shown in Fig. \ref{fig:figure1}. 

Now, a random latent embedding $z \sim p(z) := \mathcal{N}(z; 0, I)$ is introduced in cDAL $x_{\theta}$ such that $\hat{x}_{0} = x_{\theta}(x_t, t, z, I)$. Hence, the denoising model $p_{\theta}(\hat{x}_{t-1} | x_t)$ is given by:
\begin{equation}
\small
p_{\theta}(\hat{x}_{t-1} | x_t) := \int p_{\theta}(\hat{x}_{0} | x_t) q(\hat{x}_{t-1} | x_t, \hat{x}_0) d\hat{x}_0 = \int p(z) q(\hat{x}_{t-1} | x_t, \hat{x}_0 = x_{\theta}(x_t^{att}, t, z, I)) dz
\label{equation:latent_emb1}
\end{equation}
where, $p_{\theta}(\hat{x}_{0} | x_t)$ is the implicit distribution by our conditional diffusion model generator $x_{\theta}(x_t^{att}, t, z, I)$ that uses a $L$-dimensional latent variable $z$. Hence, the mapping of our conditional diffusion model label generator is $x_{\theta}(x_t^{att}, t, z, I)$. 

The predicted label $\hat{x}_0$ is not a deterministic mapping of $x_t$ as in DDPM but it is produced by the denoising model with a random latent variable $z$. This process makes the denoising distribution $p_{\theta}(\hat{x}_{t-1} | x_t)$ multimodal and hence larger step sizes could be used. The final loss to update $x_{\theta}$ (with $D$ frozen) is given by:
\begin{equation}
\small
||x_0 - x_{\theta}(x_{t}^{att}, t, z, I)||^2.
\label{equation:generator_loss}
\end{equation}
The training and sampling details of cDAL are given by Algorithms 1 and 2, respectively, which are described in the supplemental material.

\section{Experiments and Results}
\label{sec:exps}

We performed extensive experiments with our proposed cDAL algorithm on three public datasets and compared cDAL with several state-of-the-art (SOTA) segmentation methods, including SegDiff \cite{amit2021segdiff}, the best diffusion-based image segmentation model.

\subsection{Datasets}
\label{subsec: data}

\textbf{MoNuSeg dataset (2D Binary)} - 
This dataset \cite{kumar2019multi}, \cite{kumar2017dataset} consists of H\&E stained tissue images of patients with tumors of different organs. It contains 30 training and 14 held-out testing color images and corresponding binary labels. 

\noindent\textbf{CXR dataset (2D Binary)} - 
The National Library of Medicine in Maryland, USA created a standard digital chest X-ray dataset. This dataset comprises of 704 grayscale images and binary labels for the lungs, divided into 566 images for training and 138 images for testing, with a 3-fold cross-validation. 

\noindent\textbf{Hippocampus dataset (3D Multi-class)} - 
This dataset \cite{simpson2019large} is collection of 3D T1-weighted MRI images where each volume was annotated by using 2 labels for hippocampus and parts of the subiculum. The labels comprised of 3 classes: background, anterior and posterior and hence a slice by slice one-hot encoding was used to train and test the model. We divided the dataset into 130 and 65 for training and testing, with a 4-fold cross validation. 

\subsection{Experimental setup and implementation details}
\label{subsec: expsetup}
We compared cDAL with several SOTA medical image segmentation models. These include U-Net \cite{ronneberger2015u}, U-Net++ \cite{zhou2018unet++}, MedT \cite{valanarasu2021medical}, Res-UNet \cite{xiao2018weighted}, MSU-Net \cite{su2021msu}, Multi-SegCaps \cite{lalonde2018capsules}, EM-SegCaps \cite{survarachakan2020capsule}, 3D-UCaps \cite{nguyen20213d} and SegDiff \cite{amit2021segdiff}. We briefly describe SegDiff below since it is the SOTA diffusion model based algorithm for image segmentation.  
 
\noindent \textbf{SegDiff} \cite{amit2021segdiff} - Segdiff is an integration of the advanced image generation approach of diffusion models for image segmentation tasks. For the diffusion model, it uses a U-Net architecture with the input image passed as a condition through an image encoder that consists of several Residual in Residual Dense Blocks (RRDB) \cite{wang2018esrgan}. SegDiff uses 100 diffusion time-steps in its experiments.  

\noindent\textbf{Implementation details} - For cDAL, the discriminator architecture resembles the encoder part of the diffusion network, comprising of Residual blocks. Similar to other diffusion models, we utilized sinusoidal positional embeddings for time-step $t$, for both the discriminator and the diffusion model. Since the diffusion model's inference is not deterministic, following SegDiff \cite{amit2021segdiff}, we ran cDAL for 5 instance generations during the inference stage and calculated the mean segmentation map. We used PyTorch and MONAI framework for our experiments and trained our models on a NVIDIA Quadro RTX 6000 GPU.

\noindent\textbf{Evaluation metrics} - We employed three quantitative evaluation metrics. Following the literature, we used the Dice score and mIoU (mean Intersection over Union) for the CXR and MoNuSeg datasets \cite{su2021msu}\cite{amit2021segdiff} and used the Dice score, precision and recall for the Hippocampus dataset \cite{nguyen20213d}.

\subsection{Ablation study}
\label{subsec: ablation}
To assess the impact of each component in our model, we conducted an ablation study as shown in Table \ref{tab:Table1}. 

The incorporation of attention map in cDAL increases the Dice score and mIoU by up to 0.49\% and 0.79\%, respectively, on average for both the datasets. Subsequently, we identified the optimal layer in the discriminator from which we could extract the attention map. For MoNuSeg, the best layer was 32x32, while for CXR it was 16x16. One reason for this difference is that the middle layer attention maps usually contain more information about boundaries and edges (smaller-sized labels as in MoNuSeg), whereas the attention maps of the later layers typically focus on entire objects (larger labels as in CXR). Additionally, we examined the importance of the random latent embedding in our model. Without the latent embedding, the mIoU and Dice score for cDAL drops significantly with similar diffusion time-steps as our proposed cDAL, and more number of time-steps would be necessary to match the performance of cDAL.

\begin{table}[bt]
\centering
\caption{Ablation study with the MoNuSeg and chest X-ray (CXR) dataset.}
\label{tab:Table1}
\scriptsize 
\resizebox{0.95\textwidth}{!}{%
\begin{tabular}{cccc|ccc}
\hline
\multirow{2}{*}{Model} & \multicolumn{3}{c|}{MoNuSeg}                   & \multicolumn{3}{c}{CXR}                       \\ \cline{2-7} 
                       & mIoU (\%)      & Dice (\%)        & Attn. scale & mIoU (\%)      & Dice (\%)        & Attn. scale \\ \hline
cDAL w\textbackslash o Latent        & 65.95          & 79.37          & 32          & 92.50          & 96.06          & 16          \\ 
cDAL w\textbackslash o Attention     & 70.11          & 82.36          & -           & 93.27          & 96.48          & -           \\ 
cDAL                   & 70.70          & 82.77          & 16          & \textbf{94.00} & \textbf{96.87} & \textbf{16} \\ 
cDAL                   & \textbf{70.96} & \textbf{82.94} & \textbf{32} & 93.87          & 96.80          & 32          \\ 
cDAL                   & 70.38          & 82.53          & 64          & 93.68          & 96.70          & 64          \\ \hline
\end{tabular}%
}
\end{table}

\subsection{Segmentation using MoNuSeg dataset}
\label{subsec: monuseg}
Table \ref{tab:Table2} (left) presents the performance of cDAL and its comparison with several SOTA segmentation models on the MoNuSeg dataset. On a held-out test set, cDAL demonstrates a significant improvement over the other models, and an improvement of 1.96\% in mIoU and 1.35\% in Dice score over SegDiff, the current best diffusion-based segmentation model. It is worth mentioning that the notable enhancement in performance was achieved using a much lighter conditional image encoder, with 95\% less parameters than the SegDiff image encoder. Additionally, the cDAL was much faster with inference time of 1 second as it used just 4 time-steps (T=4) for training and sampling, compared to 100 time-steps (T=100) in SegDiff which takes 60 seconds inference time. Hence, our method is not computationally expensive as in inference we remove the discriminator and perform sampling with a much smaller number of steps. A qualitative comparison is provided in Fig. \ref{fig:figure2} (top row).

\begin{table}[bt]
\centering
\caption{Segmentation results on the MoNuSeg and chest X-ray (CXR) dataset. For the CXR dataset, the standard deviation across the 3-folds is indicated inside the parenthesis.}
\label{tab:Table2}
\resizebox{\textwidth}{!}{%
\tiny{
\begin{tabular}{ccc|ccc}
\hline
\multirow{2}{*}{Model}            & \multicolumn{2}{c|}{MoNuSeg}     & \multirow{2}{*}{Model}        & \multicolumn{2}{c}{CXR}                                   \\ \cline{2-3} \cline{5-6} 
                                  & mIoU (\%)      & Dice (\%)      &                               & mIoU (\%)                   & Dice (\%)                   \\ \hline
U-Net \cite{ronneberger2015u}     & 65.99          & 79.43          & U-Net \cite{ronneberger2015u} & 91.91 ($\pm$ 0.28)          & 95.75 ($\pm$ 0.17)          \\
U-Net++ \cite{zhou2018unet++}     & 66.04          & 79.49          & U-Net++ \cite{zhou2018unet++} & 92.03 ($\pm$ 0.79)          & 95.80 ($\pm$ 0.45)          \\
MedT \cite{valanarasu2021medical} & 66.17          & 79.55          & MSU-Net \cite{su2021msu}      & 92.19 ($\pm$ 0.61)          & 95.90 ($\pm$ 0.35)          \\ 
Res-UNet \cite{xiao2018weighted}  & 66.07          & 79.49          & -                             & -                           & -                           \\ 
SegDiff \cite{amit2021segdiff}    & 69.00          & 81.59          & SegDiff \cite{amit2021segdiff} & 92.33 ($\pm$ 0.69) & 95.95 ($\pm$ 0.40) \\ 
cDAL (ours)                       & \textbf{70.96} & \textbf{82.94} & cDAL (ours)                   & \textbf{93.04} ($\pm 0.97^{\textbf{*}}$) & \textbf{96.35} ($\pm 0.54^{\textbf{*}}$) \\ \hline
\end{tabular}%
}
}
\tiny{\textbf{*} indicates that the performance improvement by cDAL is statistically significant based on a t-test.}

\end{table}

\begin{figure*}[bt]
\centering
\includegraphics[width=0.92\textwidth, height=4.8cm]{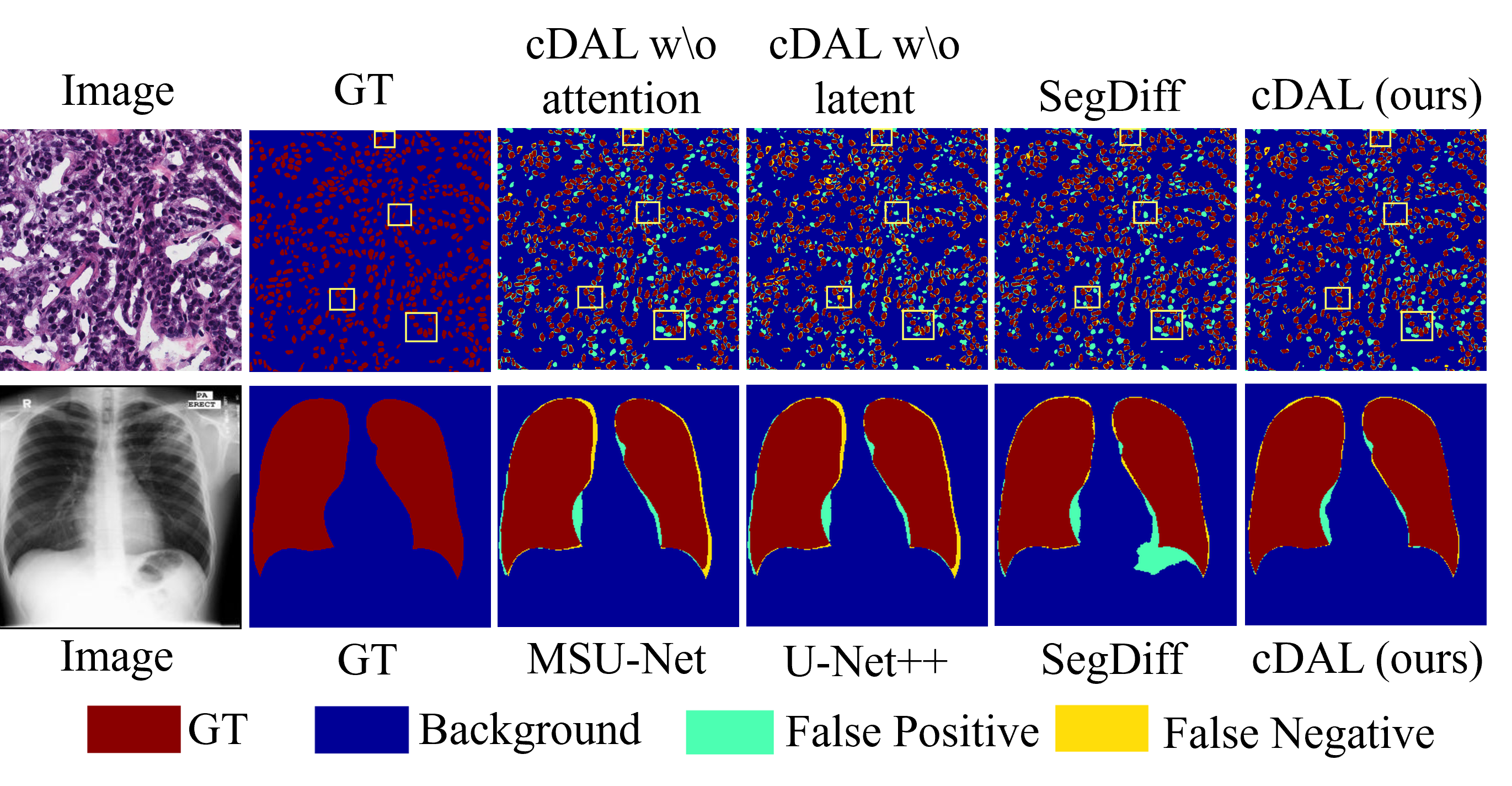}

\caption{Visualization of the Image, ground-truth (GT) and predictions with different models for MoNuSeg (top row) and CXR (bottom row) dataset. The detailed zoomed-in visual comparison is provided in the supplemental material.}
\label{fig:figure2}
\end{figure*}

\subsection{Segmentation using Chest X-ray (CXR) dataset}
\label{subsec: chestxray}
Table \ref{tab:Table2} (right) provides a comprehensive comparison between cDAL and other SOTA methods for segmentation tasks on the Chest X-ray (CXR) dataset. A 3-fold cross validation was performed to compare the performance of all models. On average, cDAL shows significant improvement in mIoU and Dice score compared to other models and an increase of 0.71\% and 0.40\% over SegDiff for mIoU and Dice, respectively. Again, cDAL is with less parameters and much faster as it achieved this performance with just 2 time-steps for training and sampling, compared to 100 in SegDiff. A qualitative comparison is provided in Fig. \ref{fig:figure2} (bottom row).

\begin{table}[bt]
\centering
\caption{Segmentation results for the Hippocampus dataset with the standard deviation across the 4-folds indicated inside the parenthesis. }
\label{tab:Table3}
\resizebox{\textwidth}{!}{%
\begin{tabular}{ccc|cc|cc}
\hline
\multirow{2}{*}{Model}                     & \multicolumn{2}{c|}{Precision (\%)}      & \multicolumn{2}{c|}{Recall (\%)}         & \multicolumn{2}{c}{Dice (\%)}           \\ \cline{2-7} 
                                           & Anterior           & Posterior          & Anterior           & Posterior          & Anterior           & Posterior          \\ \hline
Multi-SegCaps \cite{lalonde2018capsules}   & 65.65              & 60.49              & 80.76              & 84.46              & 72.42              & 70.49              \\ 
EM-SegCaps \cite{survarachakan2020capsule} & 20.01              & 34.55              & 17.51              & 19.00              & 18.67              & 24.52              \\ 
3D-UCaps \cite{nguyen20213d}               & 87.79 ($\pm$ 0.50) & 85.79 ($\pm$ 1.88) & 84.10 ($\pm$ 1.80) & 82.17 ($\pm$ 1.76) & 85.73 ($\pm$ 1.02) & 83.77 ($\pm$ 0.55) \\ 
SegDiff \cite{amit2021segdiff}             & 88.51 ($\pm$ 0.94) & 86.42 ($\pm$ 0.67) & 87.74 ($\pm$ 1.53) & 86.36 ($\pm$ 1.18) & 87.80 ($\pm$ 0.24) & 86.38 ($\pm$ 0.22) \\ 
cDAL (ours) &
  \textbf{88.76} ($\pm$ 1.15) &
  \textbf{87.43} ($\pm$ 1.50) &
  \textbf{87.85} ($\pm$ 0.79) &
  \textbf{86.72} ($\pm$ 1.25) &
  \textbf{88.13} ($\pm 0.18^{\textbf{*}}$) &
  \textbf{86.90} ($\pm 0.13^{\textbf{*}}$) \\ \hline
\end{tabular}%
}
\tiny{\textbf{*} indicates that the performance improvement by cDAL is statistically significant based on a t-test.}

\end{table}

\subsection{Segmentation using Hippocampus dataset}
\label{subsec: hippocampus}
Table \ref{tab:Table3} compares cDAL with other SOTA methods for segmentation tasks on the Hippocampus dataset. A 4-fold cross validation was performed to compare the performance of all models. On average, cDAL shows significant improvement in precision, recall and Dice score compared to other models. Compared to SegDiff, an average improvement of 0.64\%, 0.24\% and 0.43\% in precision, recall and Dice score was observed, with just 2 time-steps (T=2) for training and sampling. The visualization of the predictions is provided in the supplemental material. One limitation of our model is that it can only be applied on a 2D slice and in the future we will have a 3D version of our model.

\section{Conclusion}
\label{sec:conclusion}

In this paper, we proposed cDAL, a novel conditional diffusion model for medical image segmentation. cDAL incorporates the spatial attention from the discriminator to guide the label generation process. It also includes the random latent embedding which helped significantly reduce the number of time-steps during training and sampling. cDAL demonstrated superior results on benchmarking medical image segmentation datasets. 

\section{Disclosure of Interests}
\label{sec:disc_interest}
The authors declare that they have no competing interests in the paper.

\noindent \textbf{Acknowledgement.} Research reported in this publication was partly supported by the National Institute of Health (NIH R01HL153720).

%
%
%
\bibliographystyle{splncs04}
%

\newpage

\title{Supplemental Materials}

\titlerunning{Supplemental Materials}
%
\author{Behzad Hejrati\inst{1} \thanks{Equal contribution} 
\and
Soumyanil Banerjee \inst{1} \printfnsymbol{1} 
\and
Carri Glide-Hurst\inst{2} 
\and
Ming Dong\inst{1}
\thanks{Corresponding author}
}
\institute{Department of Computer Science, Wayne State University, MI, USA 
\email{\{b.hejrati,s.banerjee,mdong\}@wayne.edu} \and
Department of Human Oncology, University of Wisconsin-Madison, WI, USA \\
\email{glidehurst@humonc.wisc.edu}}
%
\authorrunning{B. Hejrati, S. Banerjee et. al.}

\maketitle  

\begin{algorithm}
\caption{Training algorithm for cDAL}\label{alg:Training-modified}
\begin{algorithmic}
\State \textbf{Definitions:}
\State $x_{\theta}$: Diffusion Model (generator), $D$: Discriminator, $x_0$: GT-Label, $I$: Image, $T$: Number of time steps, $A_D$: Attention Map
\For{batch of $(I, x_0)$ in Dataset}
    \State \text{Sample} $ \varepsilon \sim \mathcal{N}(0, \mathbb{I}),  z \sim \mathcal{N}(0, \mathbb{I})$
    
    \State $\beta_{\min} = 0.1$, $\beta_{\max} = 20$; $\beta_t = 1 - e^{-\beta_{\min}( \frac{1}{T}) - \frac{1}{2} (\beta_{\max} - \beta_{\min}) \frac{2t-1}{T^2}}$
    \State $\alpha_t = 1 - \beta_t$, $\bar{\alpha}_t = \prod_{s=1}^{t} \alpha_s $

    \State $t \sim \text{Uniform}(\{1, 2, \ldots, T\})$; $x_t = q(x_0, t, \epsilon )$
    
    \State $\text{Take gradient step on } D(x_t, x_{t-1}, t)$ where, $x_{t-1} = q(x_0, t-1, \epsilon)$ \& $x_{\theta}$ frozen

    \State$\hat{x}_{0} = x_{\theta}(x_{t}, t, z, I) $
    \State $\text{Take gradient step on } D(x_t, \hat{x}_{t-1}, t)$ where, $\hat{x}_{t-1} = q(\hat{x}_{0}, t-1, \epsilon)$ \& $x_{\theta}$ frozen

    \State $ A_D=(\Sigma_{i=1}^C F_i) / C $
    \State $x_{0}^{att} = x_{0} \odot  A_D $
    \State \text{Take gradient step} $\nabla_{\theta} \| x_0 - x_{\theta}(x_{t}^{att}, t, z, I)\|^2 $; $x_{t}^{att}=q({x}_{0}^{att}, t, \epsilon)$ \& $D$ frozen
\EndFor

\end{algorithmic}
\end{algorithm}

\begin{algorithm}
\caption{Sampling algorithm for cDAL}\label{alg:Inference-modified}
\begin{algorithmic}
\State \textbf{Input} $T$: number of timestemps, $I$: Images
\State $x_T \sim \mathcal{N}(0, \mathbb{I})$
\For{t $\xleftarrow[]{}$ T to 1}

    \State $\beta_{\min} = 0.1$, $\beta_{\max} = 20$; $\beta_t = 1 - e^{-\beta_{\min}( \frac{1}{T}) - \frac{1}{2} (\beta_{\max} - \beta_{\min}) \frac{2t-1}{T^2}}$
    
    \State $\alpha_t = 1 - \beta_t$, $\bar{\alpha}_t = \prod_{s=0}^{t} \alpha_s $, $\tilde{\beta}_t = \frac{1 - \bar{\alpha}_{t-1}}{1 - \bar{\alpha}_t }\beta_t $
    \State $ x_{t-1} =
    \frac{\sqrt{\alpha_t} (1 - \bar{\alpha}_{t-1})}{1 - \bar{\alpha}_t} x_t 
    +  \frac{\sqrt{\bar{\alpha}_{t-1} \beta_t}}{1 - \bar{\alpha}_t} x_\theta(x_t, t, z, I)$, where $z \sim \mathcal{N}(0, \mathbb{I})$
\EndFor
\State \textbf{return} $x_0$
\end{algorithmic}
\end{algorithm}

\begin{figure*}[ht]
\centering
\includegraphics[width=0.9\textwidth]{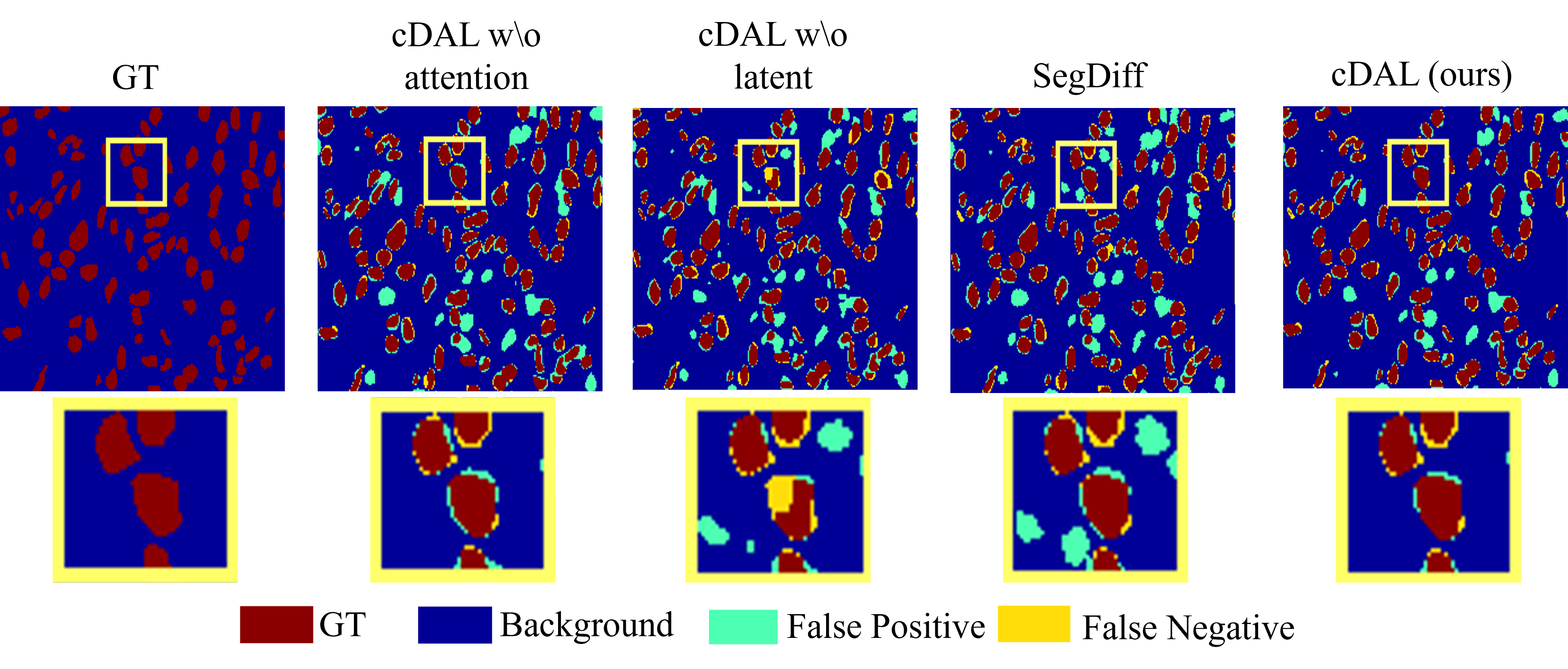}
\caption{Visualization of the Image, ground-truth (GT) and predictions with different models for MoNuSeg dataset, where the zoomed figures demonstrate the mispredictions of SegDiff, cDAL without the attention map $A_D$ and cDAL without the random latent $z$.}
\label{fig:figure1}
\end{figure*}
\vspace{-5mm}

\begin{figure*}[ht]
\centering
\includegraphics[width=0.9\textwidth]{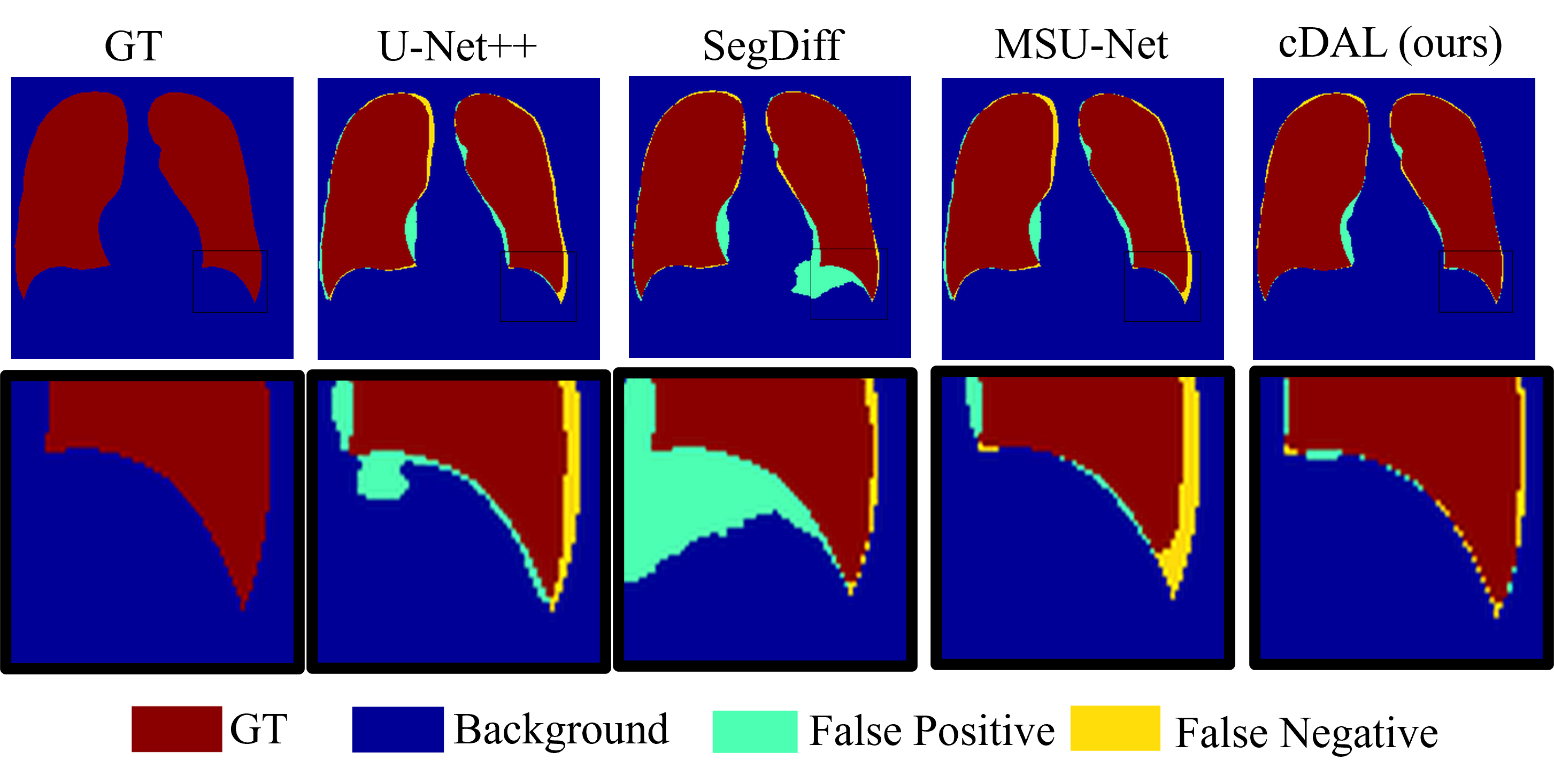}
\caption{Visualization of the Image, ground-truth (GT) and predictions with different models for the CXR dataset, where the zoomed figures demonstrate the mispredictions with other models.}
\label{fig:figure2}
\end{figure*}
\vspace{-5mm}

\begin{figure*}[ht]
\centering
\includegraphics[width=0.9\textwidth]{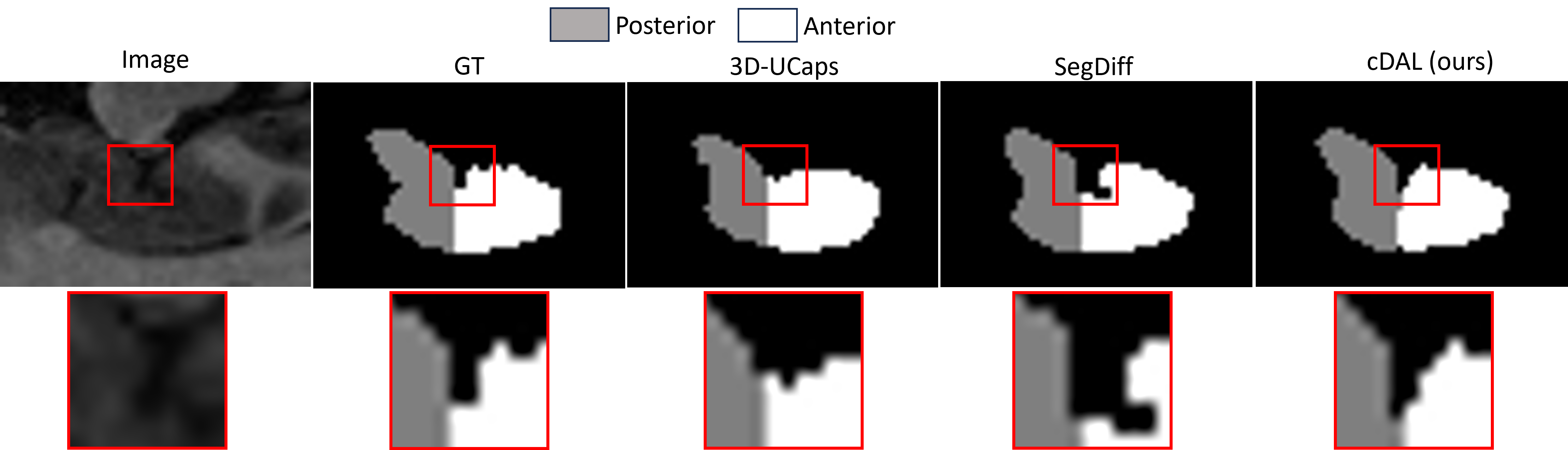}
\caption{Visualization of the Image, ground-truth (GT) and predictions with different models for Hippocampus dataset, where the zoomed figures demonstrate the mispredictions with other models.}
\label{fig:figure3}
\end{figure*}
\end{document}